\documentstyle[twocolumn,epsfig,fleqn,amstex,rotating]{mn}
\newcommand{\hi}{\mbox{H\ {\footnotesize I}}}

\newcommand{\nh}{\mbox{\scriptsize H}}
\newcommand{\nhi}{\mbox{\scriptsize H\ {\tiny I}}}
\newcommand{\nhii}{\mbox{\scriptsize H\ {\tiny II}}}

\def\etal{{et al.\ }}

\def\lsim{~\rlap{\raise 0.4ex\hbox{$<$}}{\lower 0.7ex\hbox{$\sim$}}~}
\def\gsim{~\rlap{\raise 0.4ex\hbox{$>$}}{\lower 0.7ex\hbox{$\sim$}}~}
\def\dd{{\rm d}}
\def\fion{f_{\rm ion}}

\def\ahi{\alpha_{\nhi}^{(2)}}
\def\tcbr{T_{_{\rm CBR}}}
\def\ts{T_{\rm s}}

\def\nnhii{n_{_{\nhii}}}
\def\nnhi{n_{_{\nhi}}}
\def\nnh{n_{_{\nh}}}
\def\bx{{\it \boldsymbol{x}}}

\def\mpc3{\ {\rm Mpc^{-3}}}
\def\gpc3{\ {\rm Gpc^{-3}}}

\def\hmpc{\ {\rm h^{-1}Mpc}}
\def\cm3{\ {\rm cm^{-3}}}

\begin{document}

\title[]
{The morphology of cosmological reionization by means of Minkowski Functionals}
\author[Gleser \etal]{Liron Gleser$^1$, Adi Nusser$^1$, 
Benedetta Ciardi$^2$, Vincent Desjacques$^{1,3}$\\\\
$^1$Physics Department, Technion, Haifa 32000, Israel\\
$^2$Max-Planck-Institut f\"ur Astrophysik, Karl-Schwarzschild-Stra{\ss}e 1, 85748 Garching, Germany\\
$^3$Racah Institute of Physics, Hebrew University, Jerusalem 91904, Israel}
\maketitle

\begin{abstract}
The morphology of the total gas and the neutral hydrogen (\hi)
distributions during the cosmological epoch of reionization can be
quantified with Minkowski Functionals (MFs) of isodensity surfaces.
We compute the MFs from the output of a high-resolution numerical
simulation which includes explicit treatment of transfer of UV
ionizing radiation. ``Galaxies" identified in the simulation using
semi-analytic models of galaxy formation are assumed to be the sole
sources of UV photons. The MFs of the total gas distribution are well
described by the analytic expressions derived for lognormal random
fields. The statistical properties of the diffuse \hi\ depend on the
gas distribution and on the way ionized regions propagate in the
inter-galactic medium (IGM). The deviations of the MFs of the \hi\
distribution from those of a lognormal random field are, therefore,
caused by reionization. We use the MFs to discriminate between the
various stages of reionization in the simulation. We suggest a simple
model of reionization which reproduces the MFs derived from this
simulation. Using random realizations of lognormal density fields, we
also assess the ability of MFs to distinguish between different
reionization scenarios. Our results are relevant to the analysis of
future cosmological twenty-one centimeter maps.
\end{abstract}



\section {Introduction}
\label{sec:introduction}

Current observations of the Universe probe many critical stages in its
history. Snapshots of initial density fluctuations similar to those
which have led to today's observed structure are taken by cosmic
microwave background (CMB) experiments. Galaxy redshift surveys, and
QSO spectra probe the Universe from the present epoch back to high
redshifts ($z\lsim 6$). These data have substantially tightened our
grip on the cosmological model and its fundamental parameters
(e.g. Spergel \etal 2003; Percival \etal 2002; Zehavi \etal 2002;
McDonald \etal 2005; Croft \etal 2002; Nusser \& Haehnelt 2000;
Desjacques \& Nusser 2005; Zaroubi \etal 2005; Tytler \etal 2004). But
most data leave out two important consecutive epochs. The first is the
Dark Ages starting after the formation of atomic hydrogen 380,000
years after the Big-Bang and ending with the onset of galaxy and,
possibly, QSO formation. The second is the epoch of reionization (EoR)
during which the diffuse hydrogen, the dominant primordial element in
the Universe, is reionized by radiation emitted from the emergent
luminous objects. The end of the EoR, defined when most of the diffuse
hydrogen is ionized, must have occurred before the Universe is a Gyr
old, as indicated by the amount of QSO light absorbed by neutral
hydrogen (\hi) (e.g. Gunn \& Peterson 1965; Becker \etal 2001).

The EoR leaves several observational signatures. Thomson scattering of
CMB photons off electrons released by reionization has two
consequences. First, it damps primary anisotropies by a factor that
depends on the optical depth for Thomson scattering. Second, it
introduces secondary temperature and polarization anisotropies
resulting from variations in the density and velocity fields of free
electrons in the IGM. Secondary anisotropies, expected to be observed
by ground based interferometers, will contain information on
line-of-sight integrated quantities related to the ionized gas in the
IGM. They are, however, insensitive to the details of the \hi\
component and its temporal evolution. A probe of these details can be
found in QSO spectra. Light emanating from QSOs active during the EoR
is absorbed by the intervening \hi. QSO spectra of the most distant
QSOs are therefore a direct probe of the \hi\ line-of-sight
distribution. Currently, spectra of several QSOs at sufficiently high
redshifts ($z\sim 6$) have been observed (e.g. Becker \etal
2001). However, bright QSOs are hard to detect at high redshifts, and
are therefore mainly suitable for probing the late stages of
reionization. The inference of constraints on the three-dimensional
(3D) structure of reionization from QSO spectra can be very tricky
(e.g. Nusser \etal 2002). Further, even mild \hi\ densities can
completely obscure QSO light at the relevant frequencies, preventing
an accurate estimate of the local ionized fraction. A promising probe
of reionization, especially of its initial stages, is maps of the
redshifted 21-cm line. These maps will contain 3D information on the
distribution of \hi\ and the ionized fraction as a function of
time. The 21-cm line is produced in the transition between the triplet
and singlet sub-levels of the hyperfine structure of the ground level
of neutral hydrogen atoms. This wavelength corresponds to a frequency
of 1420 MHZ and a temperature of $T_{*}=0.068{\rm \; K}$. The spin
temperature, $\ts$, is defined according to the relative population of
the triplet to the singlet sub-levels. An \hi\ region would be visible
against the CMB either in absorption if $\ts<\tcbr$ or emission if
$\ts>\tcbr$, where $\tcbr\approx 2.73(1+z)\rm \; K$ is the CMB
temperature. There are various mechanisms for raising $\ts$
significantly above $\tcbr$ during the EoR and hence a significant
cosmological 21-cm signal is expected. The prospects for measuring the
21-cm signal are excellent in view of the various radio telescopes
designed for this purpose. Despite foreground contaminations (e.g. Oh
\& Mack 2003; Di Matteo, Ciardi \& Miniati 2004), the collective data
obtained by telescopes like the Low Frequency Array (LOFAR,
e.g. Kassim \etal 2004), the Primeval Structure Telescope (PAST,
e.g. Pen, Wu \& Peterson 2004; Peterson, Pen \& Wu 2006), the Square
Kilometer Array (SKA, e.g. Carilli 2004a \& 2004b), and the Mileura
Widefield Array (MWA, Bowman, Morales \& Hewitt 2005) should, at the
very least, help us discriminate among distinct reionization
scenarios.

The statistical properties of the \hi\ distribution are dictated by
the total gas density field and the propagation of ionizing radiation
(whether UV or X-rays) emitted by galaxies and gas accreting black
holes (e.g. Ricotti \& Ostriker 2004). Therefore, 21-cm maps contain a
wealth of cosmological information. However, extracting this
information can be cumbersome. Correlation analysis of 21-cm maps
measure the power of the signal as a function of scale and may
constrain the underlying mass power spectrum. But correlations are
sensitive neither to non-Gaussian features nor to the morphology of
the large scale \hi\ distribution. Minkowski functionals (hereafter
MFs, Hadiwiger 1957), however, provide a complete set of measures of
the topology of any structure. They have been used to study the
topological and geometrical properties of isodensity surfaces of the
galaxy distribution (e.g. Mecke \etal 1994; Schmalzing \& Buchert
1997; Kerscher \etal 1997, 1998 \& 2001). MFs have also been used in
the analysis of CMB anisotropies, in particular for quantifying
non-Gaussian features (Schmalzing \& Gorski 1998; Novikov, Schmalzing
\& Mukhanov 2000; Schmalzing, Takada \& Futamase 2000; Shandarin \etal
2002; Komatsu E. \etal 2003; R\"{a}th \& Schuecker 2003; Eriksen \etal
2004). In this paper we propose the MFs of \hi\ isodensity surfaces as
measures of the morphology of reionization. We expect the MFs to be
most suitable for the analysis of 21-cm maps which will provide a
direct probe of \hi. We focus on the ability of MFs to distinguish
between the various stages of reionization and also between different
reionization scenarios, e.g. by an X-ray background, by UV ionizing
patches preferentially either in voids or dense regions. We use a
high-resolution numerical simulation (Ciardi, Ferrara \& White 2003)
in which reionization is assumed to be driven by UV photons emitted by
``galaxies" identified in the simulations using a semi-analytical
model of galaxy formation. An explicit Monte-Carlo scheme has been
used to trace the photons in the simulation box. The simulation has a
relatively small box size and it assumes only stellar type sources for
the ionizing radiation. To overcome these limitations, we use an
analytical model of reionization which produces a morphology similar
to the one obtained from the simulation. We then employ simple
reionization recipes (e.g. Benson \etal 2001) in random realizations
of lognormal density fields to explore the \hi\ morphology in
different reionization scenarios. Although this may seem simplistic,
it allows us to assess the ability of MFs to discriminate among these
scenarios.

The paper is organized as follows. In \S~\ref{sec:mfs} we review the
definition of MFs. In \S~\ref{sec:sim} we examine the morphology of
the gas and of its neutral component in a high resolution simulation
of the IGM. In \S~\ref{sec:model} we present MFs from different
reionization scenarios. We conclude with a summary and a discussion of
the results in \S~\ref{sec:discussion}.


\section{The Minkowski functionals} 
\label{sec:mfs}

\begin{figure*}
\centering
\resizebox{0.96\textwidth}{!}{\includegraphics{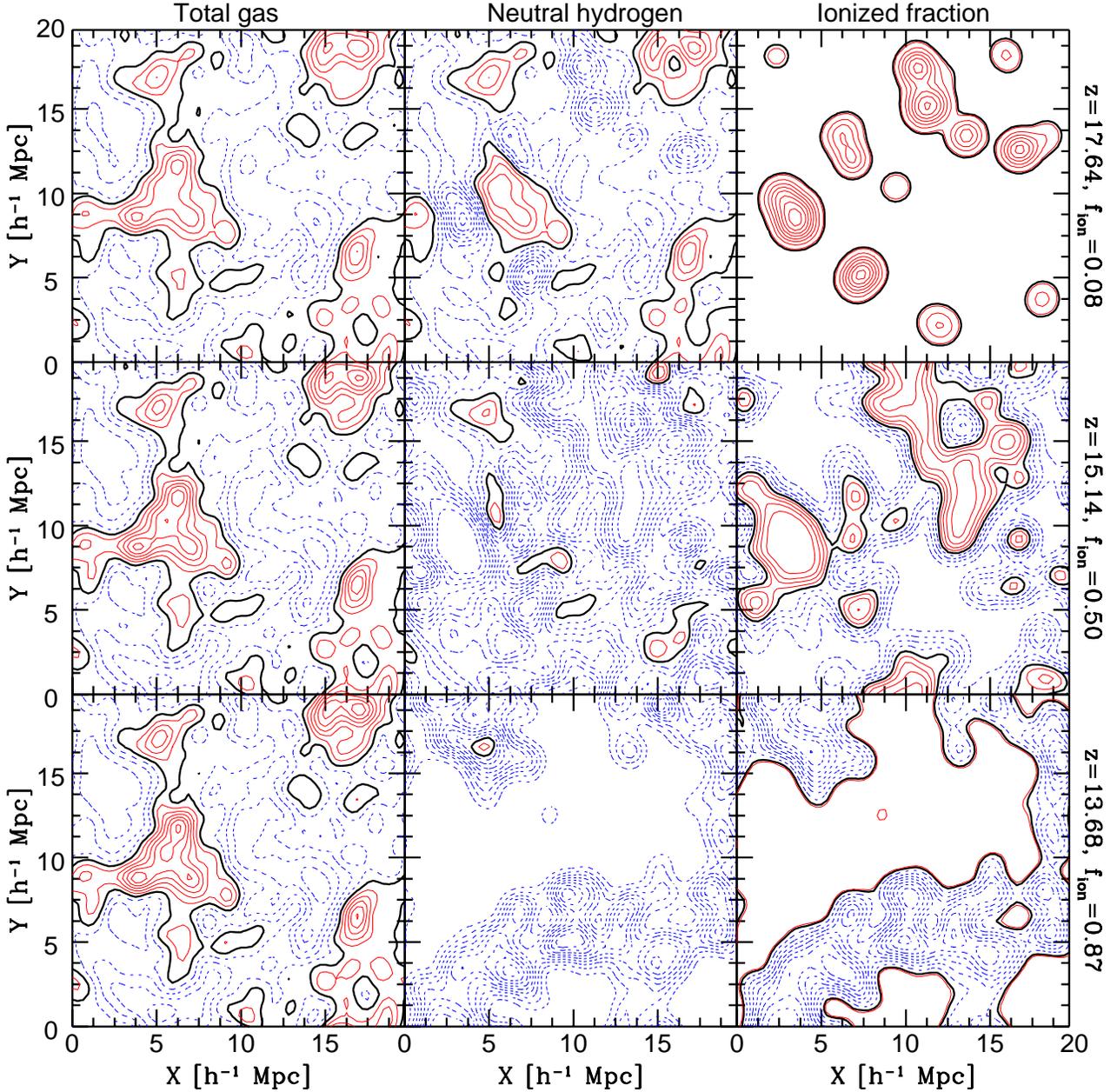}}
\caption{Fluctuations in the total gas density (left column), the \hi\
density (middle column), and the ionized fraction
$X_{_{\nhii}}=\nnhii/\nnh$ (right column) in a slice of the simulation
box at $z=17.64$ (top), $z=15.14$ (middle row), and $z=13.68$
(bottom). The mean mass weighted ionized fractions, $\fion$,
corresponding to these redshifts are indicated to the right of the
figure. In the panels showing gas and \hi\ fluctuations, the thick
solid contour indicates the mean hydrogen density ${\bar
n}_{_H}=2.66\times 10^{-7}\rm cm^{-3}$, while the thin solid and
dotted contours, respectively, represent densities above and below
${\bar n}_{_H}$. The contour spacing in these panels is $0.1{\bar
n}_{_H}$. In the panels showing $X_{_{\nhii}}$, the thick contour
shows the value of $\fion$ and the thin solid and dotted contours,
respectively, denote fractions above and below $\fion$. The contour
spacing in the panels in the right column is 0.1.}
\label{fig:ctrs}
\end{figure*}

\begin{figure*}
\centering
\resizebox{0.96\textwidth}{!}{\includegraphics{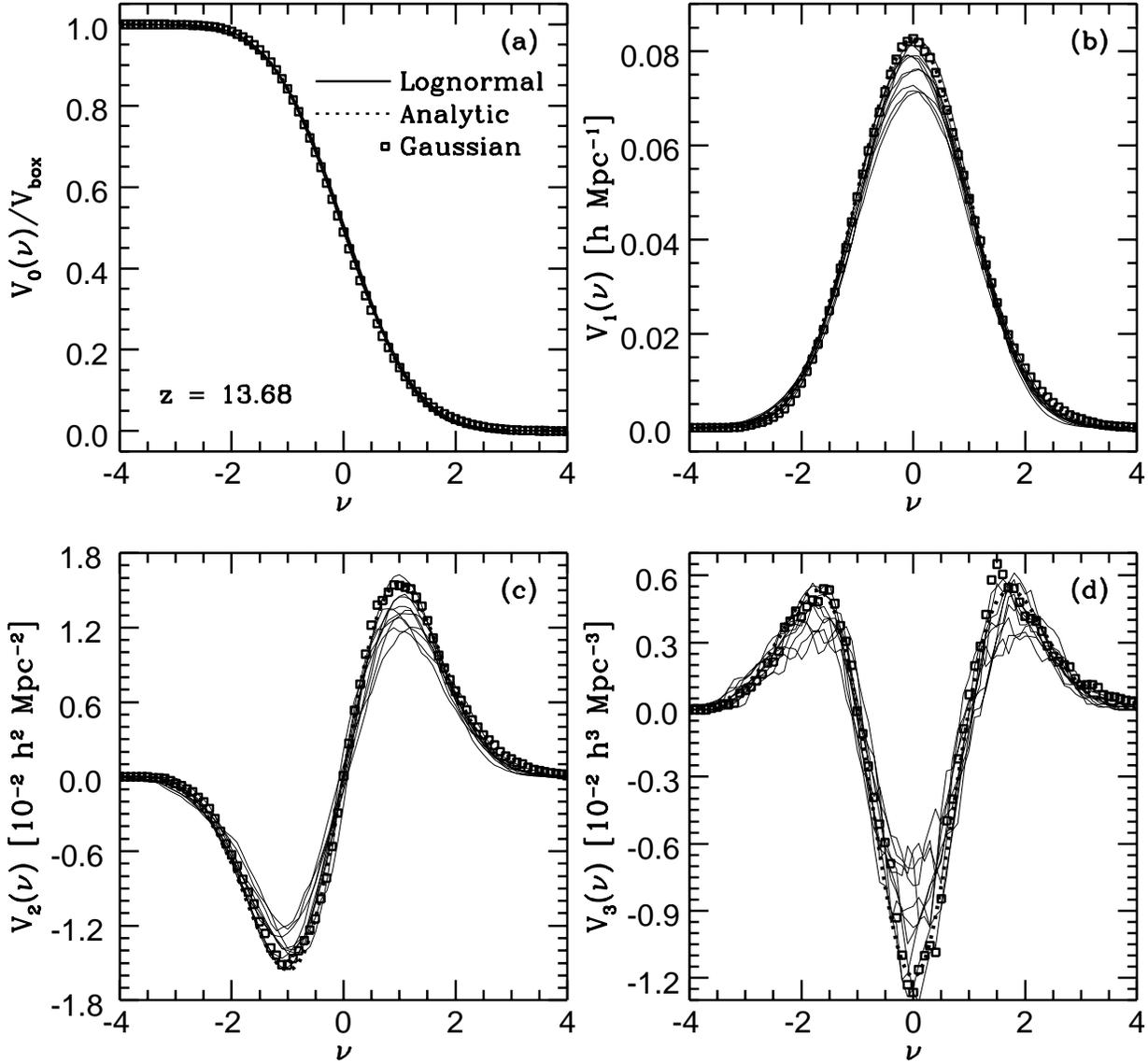}}
\caption{MFs of excursions sets defined according to a threshold in
the field $u=\ln(1+\delta_{_{\rm gas}})-\left<\ln(1+\delta_{_{\rm
gas}})\right>$, where $\delta_{_{\rm gas}}$ is the gas density
contrast smoothed with a Gaussian window of a width of $0.5\hmpc$. The
squares are obtained from the gas distribution in the simulation at
redshift $z=13.68$. The thick dotted lines represent the analytic
expressions for Gaussian fields with $\lambda$ estimated from the
field $u$. The thin solid lines are the MFs of ten Gaussian random
fields realization with the power spectrum of the $\Lambda$CDM
cosmology.}
\label{fig:lnmk}
\end{figure*}

\begin{figure*}
\centering
\resizebox{0.96\textwidth}{!}{\includegraphics{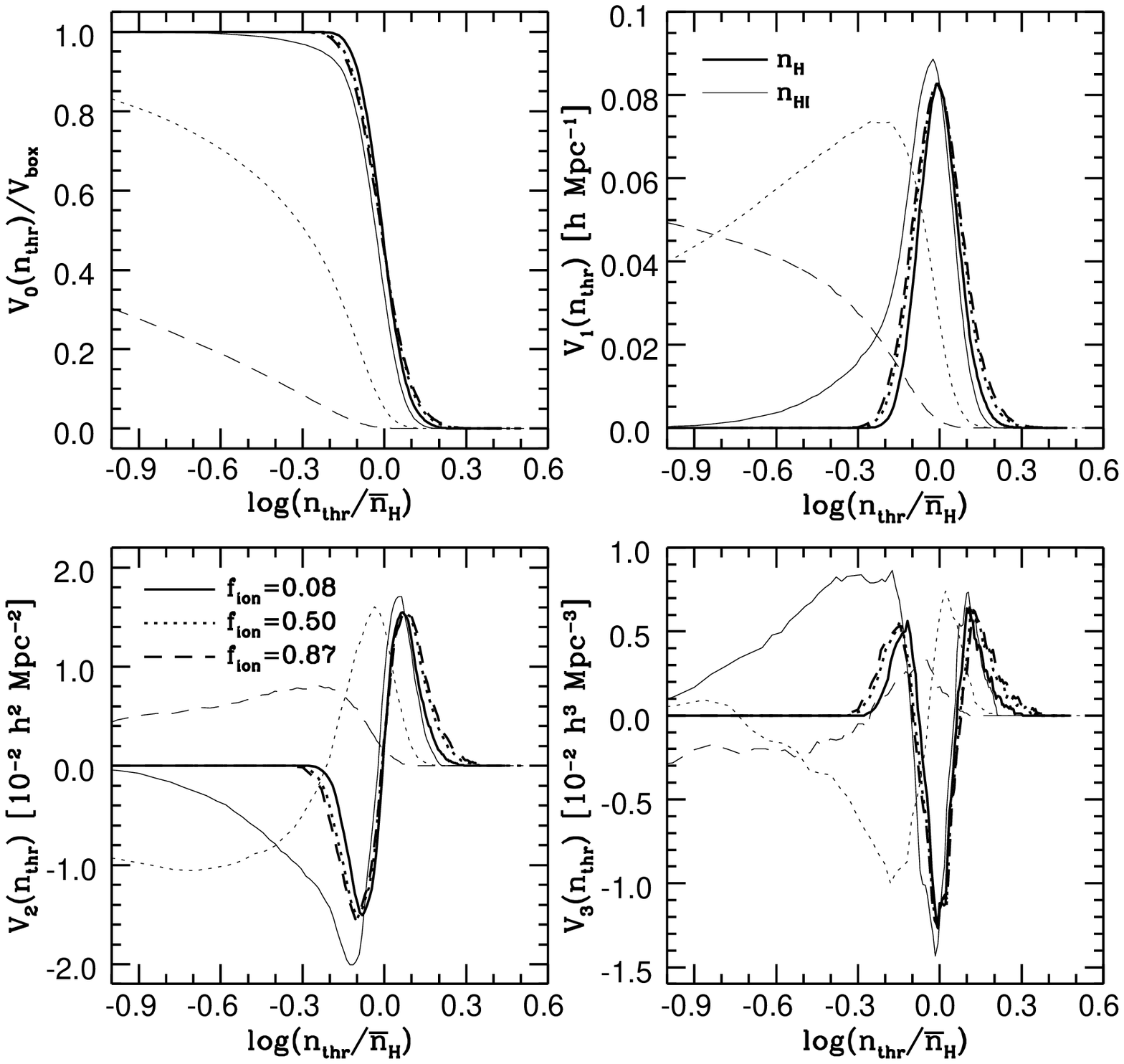}}
\caption{The MFs of gas (thick lines) and \hi\ (thin lines) isodensity
surfaces in the simulation. The gas and \hi\ density fields are
smoothed with a Gaussian window of $0.5\hmpc$ in width. Shown are
results for redshifts, $z=17.64$ (solid lines), $15.14$ (dotted
lines), and $13.68$ (dashed lines) corresponding to mass weighted mean
ionized fraction of $\fion=0.08$, $0.5$, and $0.87$, respectively.}
\label{fig:lgmk}
\end{figure*}

Let $u(\bx)$ denote a scalar function of position, $\bx$, defined in a
3D volume, $V$. Assume that $u$ has zero mean over the volume $V$ and
let $\sigma$ be the r.m.s value given by
$\sigma^2=\left<u^2\right>_{_V}$. Given a threshold value $u_{_{\rm
thr}}$ we consider the excursion set $F_\nu$ formed by all points,
$\bx$, with $u(\bx)\ge \nu\sigma$, where $\nu=u_{_{\rm
thr}}/\sigma$. The global morphology of $u(\bx)$ is then described by
calculating the MFs as a function of $u_{_{\rm thr}}$. In 3D space
there are four MFs, namely
\begin{eqnarray}
\label{eq:V0def}
V_0(\nu)&=&\frac{1}{V}\int_V\dd^3x\Theta\left(\nu\sigma-u(\bx)\right),\\
\label{eq:V1def}
V_1(\nu)&=&\frac{1}{6V}\int_{\partial F_\nu}\dd s,\\
\label{eq:V2def}
V_2(\nu)&=&\frac{1}{6\pi V}\int_{\partial F_\nu}\dd s\left[\kappa_1(\bx)+\kappa_2(\bx)\right],\\
\label{eq:V3def}
V_3(\nu)&=&\frac{1}{4\pi V}\int_{\partial F_\nu}\dd s\ \kappa_1(\bx)\kappa_2(\bx),
\end{eqnarray}
where $\Theta$ is the Heaviside step function and, $\kappa_1$ and
$\kappa_2$ are the principal curvatures (inverse of the principal
radii) at a point $\bx$ on the surface. The first MF, $V_0(\nu)$, is
the total volume of the regions with $u$ above the threshold $u_{_{\rm
thr}}=\nu \sigma$. The remaining three MFs are calculated by surface
integration over the boundary, $\partial F$, of the excursion set,
where $\dd s$ denotes the surface element on $\partial F$. The MF
$V_1$ is a measure of the surface area of the boundary, $V_2$ is the
mean curvature over the surface, and $V_3$ is the Euler
characteristic, $\chi$. Compared to a sphere, oblate (pancake-like)
surfaces are characterized by a large surface area, $V_1$, and a small
mean curvature, $V_2$. The Euler characteristic, $\chi$, can be
expressed in terms of the genus, $g$, as $\chi=2(1-g)$, where a sphere
has $g=0$ and a torus has $g=1$ (Coles, Davies \& Pearson 1996; Gott,
Weinberg \& Melott 1987). The genus can also be viewed as the number
of distinct ways an object can be sliced without being disconnected
into disjoint parts. Hence, two separate spheres have $g=-1$ since
they can be considered as an object already made of two disconnected
parts. Note that, while the Euler characteristic, $\chi$, is additive,
the genus is not so (e.g. Coles, Davies \& Peterson 1996).

For a Gaussian random field in 3D, the average MFs per unit volume have
the following analytic form (Schmalzing \& Buchert 1997)
\begin{subequations}
\begin{equation}
V_0(\nu)=\frac{1}{2}-\frac{1}{\sqrt{2\pi}}\int_0^\nu\exp\left(-\frac{x^2}{2}\right)\dd x,
\label{eq:aV0}
\end{equation}
\begin{equation}
V_1(\nu)=\frac{2}{3}\frac{\lambda}{\sqrt{2\pi}}\exp\left(-\frac{1}{2}\nu^2\right),
\label{eq:aV1}
\end{equation}
\begin{equation}
V_2(\nu)=\frac{2}{3}\frac{\lambda^2}{\sqrt{2\pi}}\nu\exp\left(-\frac{1}{2}\nu^2\right),
\label{eq:aV2}
\end{equation}
\begin{equation}
V_3(\nu)=\frac{\lambda^3}{\sqrt{2\pi}}\left(\nu^2-1\right)\exp\left(-\frac{1}{2}\nu^2\right),
\label{eq:aV3}
\end{equation}
\end{subequations}
where $\lambda=\sqrt{\sigma_1^2/6\pi\sigma^2}$,
$\sigma\equiv\left<u^2\right>^{1/2}$,
$\sigma_1\equiv\left<\left|\nabla u\right|^2\right>^{1/2}$.

Given a sampling of the field, $u(\bx)$, on a discrete cubic
lattice we use two methods to calculate the MFs (Schmalzing \& Buchert
1997). The first method is based on Koenderink invariants (Schmalzing
1996; Koenderink 1984; ter Haar Romeny \etal 1991). In this method,
the local curvatures are expressed in terms of geometric invariants
formed from the first and second derivatives of the field. The second
method is based on Crofton's formula (Crofton 1868; Schmalzing \&
Buchert 1997), which requires counting of the number of grid points,
edges, surfaces and cells within the excursion set. The appendix
\ref{apx:mfs} contains a more detailed description of these
methods. We have compared the MFs obtained from the application of the
two methods on the gas and \hi\ density fields in the
simulations. When the fields are interpolated on a $128^3$ cubic grid
using a simple cloud-in-cell algorithm, we find a difference of more
than $50\%$ between the corresponding MFs. The difference between the
results obtained with the two methods is reduced to $\lsim 10\%$ by
smoothing the density fields with a Gaussian filter of $0.5\hmpc$
width. Therefore, we work with fields smoothed with a filter of
$0.5\hmpc$. This filtering roughly corresponds to the instrumental
resolution of LOFAR (Kassim \etal 2004). We have examined the
difference between the methods using a finer grid of $256^3$ and
obtained similar differences between the methods. The Koendernick
invariants method, however, seems to be more robust and we only
present results obtained with this method.


\section{Simulations of IGM reionization} 
\label{sec:sim}

The simulation we adopt has been run by Ciardi, Ferrara \& White
(2003). This is a high-resolution numerical simulation that follows
the evolution of the dark matter and diffuse gas distribution. It is
combined to a semi-analytic model of galaxy formation which tracks the
sources of ionization (Kauffmann \etal 1999), and to the Monte-Carlo
radiative transfer code {\tt CRASH} which follows the propagation of
ionizing photons in the IGM (Ciardi \etal 2001; Maselli, Ferrara \&
Ciardi 2003). The simulation also assumes that the gas in the IGM
consists only of hydrogen. The parameters are those of a $\Lambda$CDM
``concordance'' cosmology, with $\Omega_m=0.3$,
$\Omega_{\Lambda}=0.7$, $h=0.7$, $\Omega_b=0.04$, $n=1$ and
$\sigma_8=0.9$ (Spergel \etal 2003). The smallest resolved halos have
a mass $M\simeq 10^9$~M$_\odot$, and start to form at $z\sim 20$.
Each simulation output provides a mock catalogue of galaxies which are
described, among others, by their position, star formation rate and UV
emissivity. For the specific run considered here (labeled L20 in the
original paper) reionization is completed by $z\sim 13$, and the
associated Thomson scattering optical depth is $\tau_{\rm e} 0.161$,
consistent with the result from the first year WMAP data (Spergel
\etal 2003). The total gas number density, $\nnh$, and the neutral gas
number density, $\nnhi$, are provided on a cubic grid of
$128^3$. Since we have found that MFs computed directly from the
density fields sampled on a grid without any additional smoothing are
rather noisy, we smooth the gas and \hi\ density fields with a
Gaussian filter of width of $0.5\hmpc$. All results will be presented
for the smoothed fields. Using these fields, we compute the ionized
fraction, $X_{_{\nhii}}=\nnhii/\nnh$ ($\nnhii=\nnh-\nnhi$), and the
mass weighted mean ionized fraction, $\fion=\left<\nnh
X_{_{\nhii}}\right>_V/\left<\nnh\right>_V$. Snapshots at redshift
$z=17.64$, $z=15.14$, and $z=13.68$ will be used. These correspond to
a mean ionized fraction $\fion=0.08$, $\fion=0.5$, and $\fion=0.87$,
respectively.

Fig. \ref{fig:ctrs} shows contour maps of the smoothed gas and \hi\
density fluctuations (left and middle columns, respectively) at
$z=17.64$ (top row), $z=15.14$ (middle row), and $z=13.68$ (bottom
row). The figure also shows the ionized fraction, $X_{_{\nhii}}$
(right column). In the panels to the left, the increase in the contour
levels as we go from the top to bottom is the result of the
cosmological growth of fluctuations (e.g. Peebles 1980). Non-Gaussian
features are already significant at those redshifts as indicated by
the lack of symmetry between the distribution of regions with
densities above and below the mean. A comparison between the left and
middle columns shows significant differences between the morphology of
the gas and \hi\ distribution already at the initial stages of
reionization when the mean ionized fraction is only $\fion=0.08$. At
these stages, the maps of $X_{_{\nhii}}$ (right column) indicate that
the ionized regions are mainly individual nearly spherical
patches. Recombinations play an important role at $z=17.64$ and
$15.14$ as inferred from the broad distribution of $X_{_{\nhii}}$
values (see also Ciardi \& Madau 2003).


\subsection{The morphology of the gas distribution} 
\label{sec:gas}

The evolved gas density field tends to have a lognormal probability
distribution function (e.g. Theuns \etal 1998). Consequently, we
expect the MFs of the log of the gas density field to be well
approximated by the analytic expressions for Gaussian random fields
(eqs. (\ref{eq:aV0}), (\ref{eq:aV1}), (\ref{eq:aV2}) \&
(\ref{eq:aV3})). In Fig. \ref{fig:lnmk}, we present the MFs of
$u=\ln(1+\delta_{_{\rm gas}})-\left<\ln(1+\delta_{_{\rm gas}})\right>$
where $\delta_{_{\rm gas}}$ is the smoothed gas density contrast
obtained from the simulation output at $z=13.68$. There is an
excellent match between the MFs in the simulation and the analytic
expressions. For high density thresholds, the excursion sets are
formed by individual density peaks each having an Euler characteristic
equal to that of a sphere. In that regime, $V_3$ is therefore
proportional to the number density of peaks above the threshold. Note
that, in the case of a Gaussian field, the symmetric form of $V_3$
reflects the symmetry in the statistical properties of negative and
positive peaks. Similar results were obtained at redshifts $z=17.64$
and $z=15.14$. As an estimate of the cosmic variance in the MFs
measured from the L20 simulation, we compute the scatter around the
mean MFs drawn from an ensemble of realizations of random fields with
lognormal probability distribution function (PDF). Namely, Gaussian
random realizations of the concordance $\Lambda$CDM power spectrum are
generated in a periodic box of size $L=20\hmpc$. These Gaussian fields
are mapped onto lognormal fields, which are then normalized to the
r.m.s. value of smooth density fluctuations in the simulations. The
MFs of 10 such lognormal realizations are plotted as the thin lines in
Fig. \ref{fig:lnmk}. The deviations of these MFs from the analytic
expressions (thick dotted lines) is the result of the finite size of
the box. The deviations are most pronounced in $V_3$ which depends on
$\sigma^{-3}$ (see eq. (\ref{eq:aV3})) and are smallest in $V_0$ which
does not depends on $\sigma$ at all (see eq. (\ref{eq:aV0})).


\subsection{The morphology of the \hi\ distribution} 
\label{sec:neutral}

\begin{figure}
\centering
\mbox{\epsfig{figure=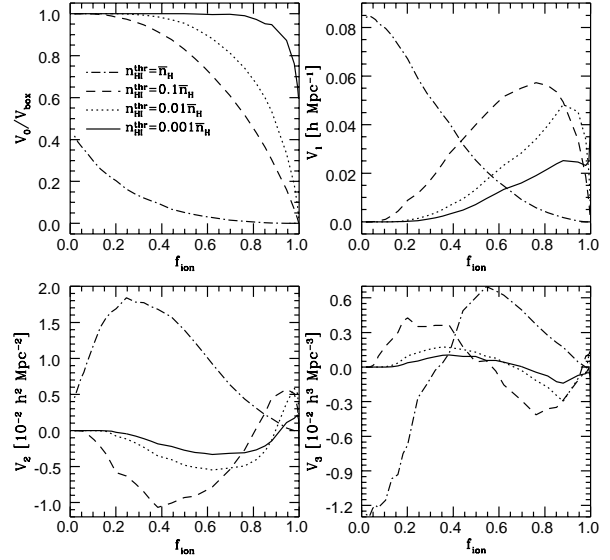,height=3.1in,width=3.2in}}
\caption{The MFs of the neutral hydrogen density as a function of the
mass weighted ionized fraction from the simulations. We present the
results for four density thresholds: $n^{\rm
thr}_{\nhi}=0.001\bar{n}_{\rm H}$ (solid lines), $n^{\rm
thr}_{\nhi}=0.01\bar{n}_{\rm H}$ (dotted lines), $n^{\rm
thr}_{\nhi}=0.1\bar{n}_{\rm H}$ (dashed lines),) and $n^{\rm
thr}_{\nhi}=\bar{n}_{\rm H}$ (dot-dashed lines), where $\bar{n}_{\rm
H}=2.66\times 10^{-7}\cm3$.}
\label{fig:shape}
\end{figure}

The reionization process in the simulation begins with the appearance
of individual ionized patches generated by the propagation of UV
ionization fronts originating from the earliest galaxies. During a
later stage, when $\fion\sim 0.5$, the geometry of ionized regions
transforms into a complex structure of tunnels and islands. Towards
the end of the EoR, UV ionizing photons form a nearly uniform
background which maintains most of the IGM at a local ionization level
determined by the balance of photo-ionization with recombinations. In
Fig. \ref{fig:lgmk}, we plot the MFs of the smoothed gas (thick lines)
and \hi\ (thin lines) number densities, for three redshifts. The
deviations of the MFs of the \hi\ distribution from those of the gas
are already significant at $z=17.64$ when the ionized fraction is only
$\fion=0.08$. Comparison between the thin (\hi) and thick (gas) lines
in the top-left panel shows that the highest density regions are all
ionized. At $z=17.64$, all regions with density $\gsim 1.42{\bar
n}_{_{\rm H}}$ are ionized together with some moderate density
regions, whereas at $z=13.68$, all the regions with mean densities and
above are ionized. Between $z=17.64$ and $z=15.14$ significant
reionization occurs in regions of densities below the mean (compare
solid and dotted thin lines). The MFs at the three redshifts are
substantially different, making the MFs a good discriminator of the
various stages of reionization. In Fig. \ref{fig:shape}, we examine
the evolution of the MFs of the \hi\ distribution as a function of the
mass weighted ionized fraction, $\fion$, for the density thresholds:
$n^{\rm thr}_{\nhi}=0.001\bar{n}_{\rm H}$ (solid lines), $n^{\rm
thr}_{\nhi}=0.01\bar{n}_{\rm H}$ (dotted lines), $n^{\rm
thr}_{\nhi}=0.1\bar{n}_{\rm H}$ (dashed lines), and $n^{\rm
thr}_{\nhi}=\bar{n}_{\rm H}$ (dot-dashed lines), where the mean gas
density is $\bar{n}_{\rm H}=2.66\times 10^{-7}\cm3$. As expected, the
volume functional, $V_0(\nu)$, decreases with $\fion$. At the initial
stages of reionization, the ionized volume of space is made of
non-overlapping patches, so that the surface area, $V_1$, is an
increasing function of $\fion$, and the integrated mean curvature
functional, $V_2(\nu)$, is negative . When the ionized fraction
$\fion$ becomes significant, the behaviour of $V_1$, $V_2$, and the
Euler characteristic $V_3$, is reversed compared to their behaviour at
low $\fion$. The reason is that both the ionized volume in the early
phase of reionization, and the neutral volume in the late stages, are
predominantly made of individual patches and, therefore, share similar
topological properties.


\section{Semi-analytic modeling of the morphology of reionization} 
\label{sec:model}

\begin{figure*}
\centering
\resizebox{0.96\textwidth}{!}{\includegraphics{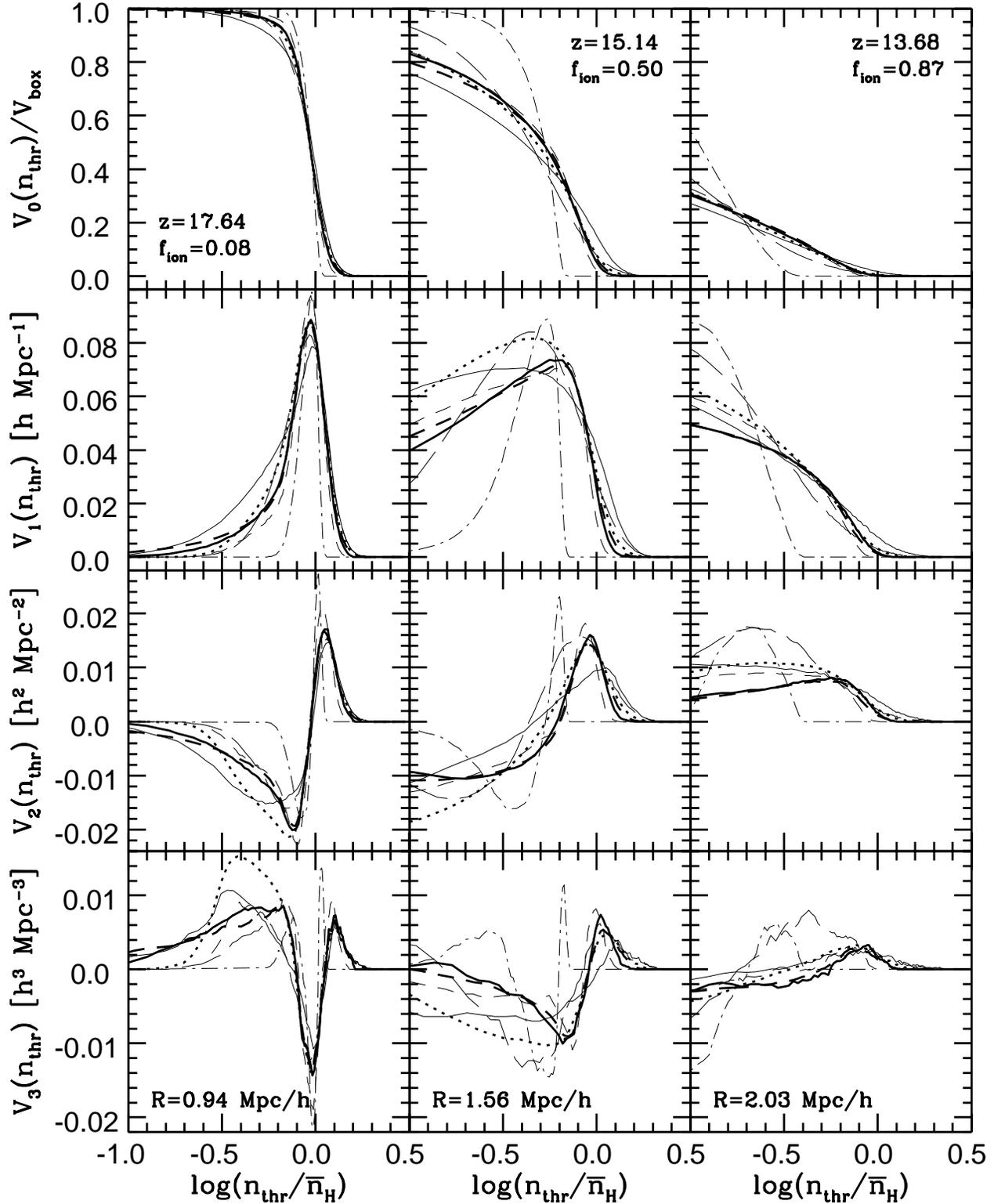}}
\caption{Comparison of the MFs of the smoothed \hi\ density field from
the simulation with those obtained from the recipes {\bf A, B, C, D,
E, \& F}. Left, middle and right columns correspond to $z=17.64$,
$z=15.14$, and $z=13.68$, respectively. The thick solid lines
represent the MFs from the \hi\ distribution in the simulation. The
thick dotted corresponds to recipe {\bf A}, thick dashed to {\bf B},
thin solid to {\bf C}, thin dashed to {\bf D}, thin long-dashed to
{\bf E}, and dash-dotted to {\bf F}. The best match to the simulation
is obtained with {\bf B} (thick dashed).}
\label{fig:rnda}
\end{figure*}

\begin{figure*}
\centering
\resizebox{0.96\textwidth}{!}{\includegraphics{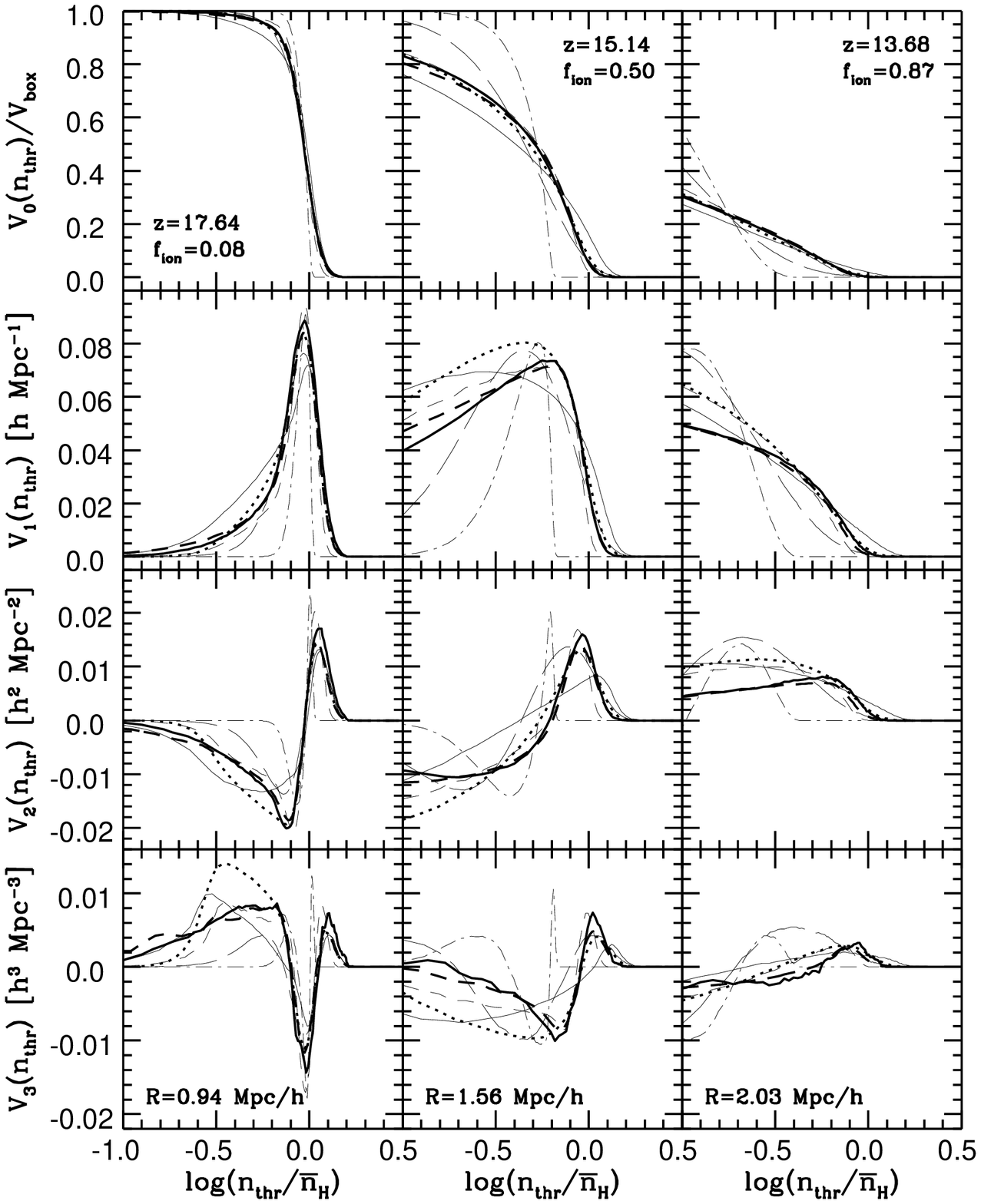}}
\caption{The same as the previous figure, but with recipes {\bf
A}--{\bf F} applied to a lognormal random density field in a cubic
box of $50\hmpc$ on the side. Here also, recipe {\bf B} (thick dashed)
yields the best match to the MFs from \hi\ field in the simulation
(thick solid).}
\label{fig:lnrc}
\end{figure*}


\subsection{Modeling the morphology in the simulation} 
\label{sec:model_sim}

Our goal here is twofold. First, we will attempt to reproduce the MFs
of the \hi\ distribution in the simulation using simple recipes that
do not require explicit treatment of radiative transfer. Second, given
these recipes we will assess the extent to which MFs can discriminate
between reionization scenarios other than those adopted in the
particular simulation used here. In the recipes {\bf A}, {\bf B}, {\bf
C}, and {\bf D}, that we describe below, the \hi\ distribution is
obtained by overlaying the gas distribution with spherical bubbles
devoid of neutral gas, while the surrounding gas is kept neutral. In
the recipes {\bf E} and {\bf F} the IGM in the simulation box is
ionized cell by cell as explained below. All the models assume that a
grid cell is either completely ionized or neutral. Clearly, this is an
oversimplification especially in the early stages of reionization (see
column to the right in Fig. \ref{fig:ctrs}). Yet we will see that one
of these recipes indeed reproduces the MFs measured in the
simulations. We now summarize the different recipes.
\begin{itemize}
\item {\bf A}: spherical bubbles are placed randomly in the
simulation. All bubbles have the same radius, $R$, independent of
local gas density.
\item {\bf B}: the same as {\bf A}, but with bubble radius, $r$,
depending on the local density. We have tried several functional
forms, and found that the following parametrization gives 
the best match to the MFs of the simulation,
\begin{equation}
r(\bx)=\left\{
\begin{array}{ll}
R+6R\left[1-\left(1+\delta(\bx)\right)^{-1/3}\right] & 1+\delta\ge\left(6/7\right)^3 \\
0 & 1+\delta<\left(6/7\right)^3
\end{array}\right.,
\label{eq:scale}
\end{equation}
where $\delta(\bx)$ is the smoothed density contrast at the
position $\bx$, and $R$ is a constant.
\item {\bf C}: bubbles of fixed radius $R$, are added one by one,
where a newly added bubble is centered on the highest density grid
point not encompassed by previous bubbles.
\item {\bf D}: the same as {\bf C} but with new bubbles centered at
lowest density point.
\item {\bf E}: grid cells are ionized according to their density,
starting from the cell with the highest gas density.
\item {\bf F}: the same as {\bf E} but starting from the lowest
density cell.
\end{itemize}

We apply each of the recipes described above to the gas distribution
in the L20 simulation. To facilitate the comparison with the MFs
derived from the \hi\ distribution in the simulation, the number of
bubbles in {\bf A}--{\bf D}, and the number of ionized cells in {\bf
E} \& {\bf F} are tuned so that the mean mass weighted ionized
fraction matches the value measured in the simulation at a given
redshift. In Fig. \ref{fig:rnda}, we compare the MFs obtained with
these recipes to those of the actual \hi\ distribution in the
simulation. Left, middle, and right columns show, respectively,
results for $z=17.64$, 15.14, and 13.88. In each of the models {\bf
A}--{\bf D}, the parameter $R$ is tuned to the value giving the best
match to the curve of $V_0$ obtained from the simulation. We have
found that this value mainly depends on redshift: $R=0.94$, 1.56, and
$2.03\hmpc$ comoving for $z=17.64$, 15.14, and 13.88,
respectively. The figure shows results for these values of $R$
only. Recipe {\bf B} (thick dashed lines) yields the best fit to the
MFs of the actual \hi\ distribution in simulation (think solid
lines). In this model, the radius of bubbles in regions with
overdensities $\lsim 0.63$ is practically zero. Therefore, these
regions are only ionized by sources in nearby denser regions. This
reflects the relative lack of ionizing sources (galaxies) in the voids
of the simulation volume.

The numerical simulation we use has a relatively small box ($20\hmpc$)
so that cosmic variance may introduce large uncertainties in the
estimates of the MFs (see thin lines in Fig. \ref{fig:lnmk}).
Furthermore, it is assumed in this simulation that reionization is
caused solely by UV radiation. The amount of UV photons emitted by a
``galaxy" in the simulation is sensitive to the adopted parameters of
the semi-analytic galaxy formation model. The distribution of the
``galaxies", which greatly affects the development of the early
stages, also depends on the galaxy formation model. Exploring the
parameter space of the galaxy formation model and all possible
scenarios with large volume simulations is currently
unfeasible. Nonetheless, Figs. \ref{fig:rnda} demonstrate that the MFs
can distinguish between the different reionization scenarios examined
here.

To reduce the effect of cosmic variance in the MFs we apply the
recipes {\bf A}--{\bf F} to lognormal random fields in a cubic box of
$50\hmpc$ on the side (more than 15 times the volume of the L20
simulation). The corresponding MFs are shown in
Fig. \ref{fig:lnrc}. The larger box size significantly reduces the
noise in the MFs, especially for $V_3$. In general, there are only
minor differences between Figs. \ref{fig:rnda} and \ref{fig:lnrc},
indicating that the lognormal approximation yields results in
agreement with the simulation. This is expected since the morphology
of the total gas density is well approximated by the MFs of a
lognormal field. Therefore, the morphology of reionization can also be
modeled using lognormal fields without resorting to detailed
simulations.


\subsection{Morphology in an X-ray pre-reionization scenario}
\label{sec:xray}

\begin{figure}
\centering
\mbox{\epsfig{figure=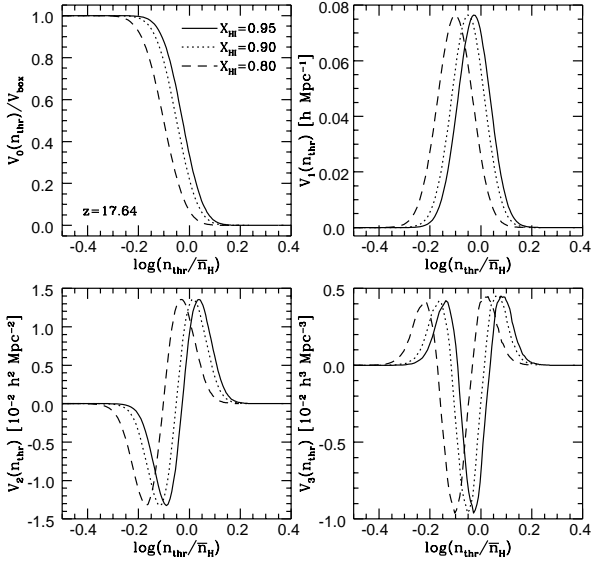,height=3.1in,width=3.2in}}
\caption{MFs of excursions sets obtained from the \hi\ distribution in
an X-ray pre-reionization scenarios. Plotted are curves for three
values of $X_{_{\nhi}}$ at mean gas density, as indicated in the
figure. All curves are for $z=17.64$.}
\label{fig:xray}
\end{figure}

\begin{figure}
\centering
\mbox{\epsfig{figure=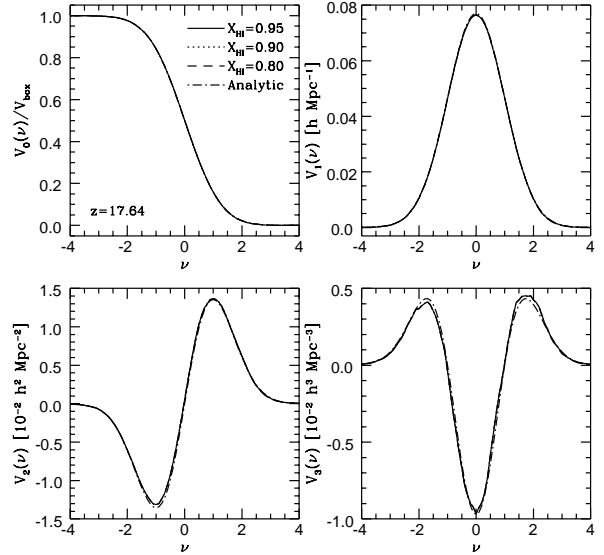,height=3.1in,width=3.2in}}
\caption{The MFs from the previous figure plotted against threshold in
$\nu=\left[\ln(\nnhi)-\left<\ln(\nnhi)\right>\right]/\sigma$ where
$\sigma^2=
\left<\left[\ln(\nnhi)-\left<\ln(\nnhi)\right>\right]^2\right>$. The
analytic MFs for Gaussian random field are shown as the dash-dotted
lines.}
\label{fig:lnx}
\end{figure}

Reionization by UV sources can be preceded by a pre-reionization epoch
in which the gas is partially ionized by an X-ray background produced
by accretion onto mini black holes (Ricotti \& Ostriker 2004). The
mean free path of X-ray photons is much larger than both the
correlation length and the mean separation of the sources. Therefore,
in this scenario, X-ray photons act as a uniform ionizing background
and the \hi\ fraction, {\bf $X_{_{\nhi}}=\nnhi/\nnh$}, at a point with
density contrast $\delta$, is determined by the equation for
photo-ionization equilibrium,
\begin{equation}
\Gamma X_{_{\nhi}}=\ahi{\bar n}(1+\delta)(1-X_{_{\nhi}})^2 \; , 
\label{eq:photo}
\end{equation}
where $\Gamma$ is the X-ray photo-ionization rate per hydrogen atom,
$\ahi$ is the recombination cross section to the second excited atomic
level, $\bar n$ is the mean density of total (ionized plus neutral)
hydrogen, and $X_{_{\nhi}}$ is the \hi\ fraction. According to Ricotti
\& Ostriker (2004), the mean \hi\ fraction should be maintained at a
level $X_{_{\nhi}}\sim 0.8-0.9$ in the redshift range $20\lsim z\lsim
40$ in order to yield a large optical depth $\tau_{\rm e}\sim
0.1$. This level of reionization is ideal for producing a significant
21-cm signal (Nusser 2005a; Kuhlen, Madau \& Montgomery 2006). In
order to model the MFs of \hi\ isodensity surfaces in this scenario,
we use the lognormal gas density fields which have been computed from
Gaussian random fields in a box of $50\hmpc$ on the side. We fix the
ratio $\ahi{\bar n}/\Gamma$ by assuming that $X_{_{\nhi}}$ is given at
mean density, i.e. at $\delta=0$. Then, for any $\delta$,
$X_{_{\nhi}}$ can be easily found using eq. (\ref{eq:photo}). In
Fig.~\ref{fig:xray} we present the MFs of excursions sets obtained
from X-ray background \hi\ fields at redshift $z=17.64$ for values of
the \hi\ fraction at mean density of $X_{_{\nhi}}=0.95$ (solid lines),
0.9 (dotted lines), and 0.8 (dashed lines). Changing the \hi\ fraction
shifts the MFs horizontally, but does not affect their overall
shape. This is not surprising since all these models have similar
topologies. In Fig.~\ref{fig:lnx}, we plot the MFs as functions of
excursion sets obtained from the field $\ln(\nnhi)$. Also plotted, as
the dash-dotted lines, are the analytic expressions for Gaussian
random field. The figure demonstrates that for the low levels of
ionization obtained in X-ray pre-reionization, the MFs of the \hi\
distribution are close to those of a lognormal field which also
describes the morphology of the gas distribution in our case.


\section{Discussion}
\label{sec:discussion}

Among all possible measures of global morphology, only Minkowski
Functionals (MFs) are invariant under rotational and translational
transformations (Hadwiger 1957). But can the MFs distinguish between
various stages of EoR or between different reionization scenarios? We
have assessed this question using a high-resolution numerical
simulation in which reionization is driven by UV photons emitted from
stellar type sources and by employing simple recipes to describe
different reionization scenarios. We have found that the MFs of
isodensity surfaces of the gas distribution in the simulation are well
described by those of a field having a lognormal distribution
function. In a UV reionization scenario, the deviations from the
lognormal MFs are sensitive to the distribution of ionized
regions. The MFs of the \hi\ distribution depend on a multitude of
factors among which are: the gas density field and its correlation
with the distribution of ionizing sources, the abundance of sources,
and the intensity and type (UV versus x-rays) of the emitted
radiation.  Reionization by UV begins with individual patches
appearing around sources associated with the most massive and rarest
haloes (e.g. Benson \etal 2005). The topology of reionization in these
early stages is substantially different from that of later stages when
ionized regions form a complex network of tunnels in addition to
isolated patched. This difference in topology is reflected in the
shapes of the MFs. X-rays have a substantially larger mean free path
than UV photons. Therefore, x-rays generated by an early generation of
black holes that tend to form a uniform background that may partially
ionize the IGM. We have considered the MFs of the \hi\ distribution in
an epoch of pre-reionization caused by an x-ray background. Using a
semi-analytic modeling of this epoch, we find similar \hi\ and gas
topologies. This is due to the low local ionized fractions obtained in
these models.

Despite the complexity of the reionization process as given by the
simulation (cf. Ciardi, Ferrara \& White 2003), its topology can be
described using simple semi-analytic recipes that do not involve a
detailed treatment of radiative transfer. Application of some of these
recipes on lognormal random fields yield a good match to the MFs
obtained from the simulation. However, one may envisage a variety of
alternative UV reionization scenarios (e.g. Miralda-Escud\'e, Haehnelt
\& Rees 2000), depending on the details of galaxy formation and the
dark matter distribution. It therefore remains to be confirmed whether
semi-analytic recipes can actually match the MFs obtained from
detailed simulations of alternative scenarios corresponding to
physical assumptions different from those adopted in the simulation we
have used in this paper.

In the near future, the next generation of radio telescopes will
provide maps of twenty-one centimeter (21-cm) line from neutral IGM
and probe the three dimensional distribution of \hi\ during the
EoR. It will be then possible to use MFs to study the morphology of
the reionization process as described in this work. However, the
observed 21-cm maps will be affected by redshift distortions
(e.g. Ali, Bharadawaj \& Pandey 2005; Barkana \& Loeb 2005; Nusser
2005b), instrumental noise, and foreground contamination. The
usefulness of MFs derived from 21-cm maps which include all these
effects has yet to be demonstrated. This will be the subject of future
work in which the modified methodology of Benson \etal (2001, 2005)
for studying reionization will be implemented in a high resolution
N-body simulation of a large box ($\sim 140\hmpc$ on the side).

The background density parameters and the value of $\sigma_8$ of the
simulation used in this paper are consistent with the first year WMAP
data (Spergel \etal 2003). Although the third year WMAP data give
similar values for the density parameters as the first year data, it
yields $\sigma_8=0.74^{+0.05}_{-0.06}$ and $\tau_{\rm
e}=0.09^{+0.03}_{-0.03}$ (Spergel \etal 2006). The difference in the
values for $\sigma_8 $ and $\tau_{\rm e}$ can amount to substantial
changes in the reionization history (Benson \etal 2005). This work has
been completed before the the third year data has been published and
therefore simulations have been run with cosmological parameters
inferred from the first year WMAP data. Nevertheless, our aim at this
stage is to illustrate the use of MFs for studying the history of
cosmic reionization. Implications of MFs of 21-cm maps generated from
a suite of simulations will be presented elsewhere.

Auto-correlations estimated from 21-cm maps have also been proposed as
a statistical measure of \hi\ distribution. The advantage of
auto-correlations is that redshift distortions, noise, and foreground
contamination can probably be easily incorporated in or removed from
them, while these effects might play a more important or ambiguous
role in MFs of 21-cm maps. Nevertheless, as auto-correlations are
insensitive to morphology (e.g. a Gaussian random field and another
highly non-Gaussian one, can have identical auto-correlation
functions), MFs are better discriminators between different
reionization scenarios. Therefore, an exploration of MFs of 21-cm maps
is worthwhile.


\section*{Acknowledgments}

AN wishes to thank the Max-Planck-Institut f\"ur Astrophysik in
Garching for its generosity and support. AN \& LG acknowledge a useful
discussion with T. Buchert. VD acknowledges the support of a Golda
Meir fellowship.



\appendix

\section{Numerical calculation of the Minkowski functionals} 
\label{apx:mfs}

Schmalzing \& Buchert (1997) used an integral geometry approach and
developed two numerical methods to investigate the morphology of a
sampled density field on a cubic lattice grid. Using the techniques
pioneered by Koenderink (Koenderink 1984; ter Haar Romeny \etal 1991) 
in two-dimensions, Schmalzing (1996) expressed the local curvatures 
in term of geometric invariants formed from first and second
derivatives. These invariants are known as the Koenderink invariants.
\begin{subequations}
\begin{equation}
\kappa_1+\kappa_2=\frac{\epsilon_{ijm}\epsilon_{klm}u_{,i}u_{,jk}u_{,l}}{(u_{,n}u_{,n})^{3/2}},
\label{eq:kdr2}
\end{equation}
\begin{equation}
\kappa_1\kappa_2=\frac{\epsilon_{ijk}\epsilon_{lmn}u_{,i}u_{,l}u_{,jm}u_{,kn}}{2(u_{,p}u_{,p})^2},
\label{eq:kdr3}
\end{equation}
\end{subequations}
where $\kappa_1$ and $\kappa_2$ are the principal curvatures, $u({\bf
x})$ is a three-dimensional density field in a box of a volume $V$,
$u_{,i}$ and $u_{,ij}$ are the first and second spatial derivatives of
the field $u(\bx)$ respectively in the $i$ and $j$ directions, and
$\epsilon_{ijk}$ is the Levi-Civita tensor.

Since one can replace the surface integration in eqs.
(\ref{eq:V1def}), (\ref{eq:V2def}) \& (\ref{eq:V3def}) with a spatial
mean over the whole volume, the MFs become
\begin{subequations}
\begin{equation}
V_1(\nu)=\frac{1}{6V}\int_V\dd^3x\delta(\nu\sigma-u(\bx))\left|\nabla u(\bx)\right|,
\label{eq:V1kdr}
\end{equation}
\begin{equation}
V_2(\nu)=\frac{1}{6\pi V}\int_V\dd^3x\delta(\nu\sigma-u(\bx))\left|\nabla u(\bx)\right|\left[\kappa_1(\bx)+\kappa_2(\bx)\right],
\label{eq:V2kdr}
\end{equation}
\begin{equation}
V_3(\nu)=\frac{1}{4\pi V}\int_V\dd^3x\delta(\nu\sigma-u(\bx))\left|\nabla u(\bx)\right|\kappa_1(\bx)\kappa_2(\bx),
\label{eq:V3kdr}
\end{equation}
\end{subequations}
where $\nu$ is the density threshold divided by the field
r.m.s. value, $\sigma$, and $\delta$ is a delta function.

The second method is based on Crofton's formula (Crofton 1868) which
requires counting of the number of grid points, edges, surfaces and
cells within the excursion set. The MFs in three-dimensions are given
by
\begin{subequations}
\begin{equation}
V_0(K)=\frac{1}{L}N_3(K),
\label{eq:V0cft}
\end{equation}
\begin{equation}
V_1(K)=\frac{2}{9}\frac{1}{aL}\left[-3N_3(K)+N_2(K)\right],
\label{eq:V1cft}
\end{equation}
\begin{equation}
V_2(K)=\frac{2}{9}\frac{1}{a^2L}\left[3N_3(K)-2N_2(K)+N_1(K)\right],
\label{eq:V2cft}
\end{equation}
\begin{equation}
V_3(K)=\frac{1}{a^3L}\left[-N_3(K)+N_2(K)-N_1(K)+N_0(K)\right],
\label{eq:V3cft}
\end{equation}
\end{subequations}
where the body $K$ is the excursion set of a homogeneous and isotropic
random field sampled at $L$ points of a cubic lattice of spacing
$a$. $N_3(K)$ gives the number of cubes within the body, $N_2(K)$ is
the number of faces and $N_1(K)$ is the number of edges of the lattice
cubes which belong to the body, while $N_0(K)$ gives the number of
lattice points contained in $K$.

In http://physics.technion.ac.il/\~{}lirong/mfs.tar.gz one can find
the {\tt MFs} software package for MFs calculation. This package
includes the Koenderink invariants method, the Crofton's formula
method and the analytical calculation for Gaussian random field.


\end{document}